\documentclass[final,5p,times,twocolumn]{elsarticle}

\usepackage{amssymb}
\journal{Physica E}

\def\Ham{\mathcal{H}}
\def\kagome{kagom\'{e}}
\def\Nd{N_{\rm d}}

\begin{document}

\begin{frontmatter}

%% Title, authors and addresses

%% use the tnoteref command within \title for footnotes;
%% use the tnotetext command for the associated footnote;
%% use the fnref command within \author or \address for footnotes;
%% use the fntext command for the associated footnote;
%% use the corref command within \author for corresponding author footnotes;
%% use the cortext command for the associated footnote;
%% use the ead command for the email address,
%% and the form \ead[url] for the home page:
%%
%% \title{Title\tnoteref{label1}}
%% \tnotetext[label1]{}
%% \author{Name\corref{cor1}\fnref{label2}}
%% \ead{email address}
%% \ead[url]{home page}
%% \fntext[label2]{}
%% \cortext[cor1]{}
%% \address{Address\fnref{label3}}
%% \fntext[label3]{}

\title{Quantum Field Induced Orderings in Fully Frustrated Ising Spin Systems}

%% use optional labels to link authors explicitly to addresses:
%% \author[label1,label2]{<author name>}
%% \address[label1]{<address>}
%% \address[label2]{<address>}

\author[label1]{Shu Tanaka}
\author[label2]{Masaki Hirano}
\author[label2,label3]{Seiji Miyashita}

\address[label1]{Institute for Solid State Physics, University of Tokyo, 5-1-5 Kashiwanoha, Kashiwa-shi, Chiba, 277-8581, Japan}
\address[label2]{Department of Physics, University of Tokyo, 7-3-1, Hongo, Bunkyo-ku, Tokyo, 113-0033, Japan}
\address[label3]{CREST, JST, 4-1-8, Honcho Kawaguchi, Saitama, 332-0012, Japan}

\begin{abstract}
%% Text of abstract
We study ordering mechanism which is induced by a quantum
 fluctuation in fully frustrated Ising spin systems. 
 Since there are many degenerated states in frustrated systems, ``order by
 thermal disorder'' often takes place due to a kind of entropy effect. 
 To consider ``order by quantum disorder'' in fully frustrated Ising spin
 systems, 
 we apply transverse field as a quantum fluctuation.
 In triangular antiferromagnetic Ising spin system, there exists a ferromagnetic correlation in each sublattice.
 The sublattice correlation at zero temperature is enlarged due to transverse field.
 The quantum fluctuation enhances the solid order at zero temperature. 
 This is an example of quantum field induced ordering in fully frustrated systems. 
 We also study a case in which the transverse field induces a reentrant
 behavior as another type of order by quantum disorder, and compare correspondent cases in the classical systems.
\end{abstract}

\begin{keyword}
%% keywords here, in the form: keyword \sep keyword
transverse Ising model \sep frustration \sep order by disorder \sep
 reentrant phase transition
%% MSC codes here, in the form: \MSC code \sep code
%% or \MSC[2008] code \sep code (2000 is the default)

\end{keyword}

\end{frontmatter}

%%
%% Start line numbering here if you want
%%
% \linenumbers

%% main text
\section{Introduction}
\label{sec:introduction}

Orderings in frustrated systems are very interesting topics and have
been studied by a number of researchers.
Because frustrated systems, such as triangular, \kagome, and pyrochlore
antiferromagnets, have many degenerated states, entropy plays an important role in ordering mechanism.
In unfrustrated model, such as ferromagnetic Ising model,
ferromagnetic order destroys by increasing temperature.
On the other hand, in geometrical frustrated systems, orderings due to thermal
fluctuation often occurs, which is called ``order by disorder''~\cite{Villain-1980,Chalker-1992,Huse-1992,Reimers-1993,Moessner-2001,Tanaka-2007-1}.
For example, Villain et al. have studied homogeneous fully frustrated
Ising spin system which is the so-called ``Villain model'' and concluded
existence ordering owing to thermal fluctuation in this system~\cite{Villain-1980}.
Other example is reentrant phase transition where the order parameter
shows non-monotonic behavior as a function of temperature~\cite{Syozi-1972,Miyashita-1983,Kitatani-1986,Miyashita-2001,Tanaka-2005,Miyashita-2007-1}.
Generally speaking, because frustration makes peculiar density of
states, thermal fluctuation induced ordering can occur.
Other type of fluctuation is quantum fluctuation.
It is well-known that quantum fluctuation also makes orderings as well
as thermal fluctuation~\cite{Chubukov-1992,Moessner-2000}.
For example, Moessner, Sondhi, and Chandra have studied quantum dimer
model~\cite{Moessner-2000}.
This model corresponds to classical dimer covering problem and Ising
model with transverse field on its dual lattice by controlling the
parameter of the hopping dimer and the on-site potential.
In quantum dimer model, there are many ordering structure due to quantum fluctuation.

Motivation of our study is to clarify the effect of the quantum fluctuation comparing with the effect of the thermal fluctuation.
We study the sublattice correlation function at the ground state as a
function of transverse field on triangular Ising antiferromagnets.
We also consider the quantum fluctuation induced reentrant behavior on frustrated decorated bond system.

\section{Model}
\label{}

In this paper, we study the ground state properties of geometric fully
frustrated Ising model with transverse field by exact diagonalization method
for small system and by power method for relatively large system.
We consider transverse Ising model on triangular antiferromagnets and on
decorated bonds where thermal ({\it i.e.} classical) reentrant phase transition occurs.

\subsection{Triangular Ising Antiferromagnets}
\label{}

Most well-known geometric fully frustrated Ising model is triangular
Ising antiferromagnets.
Because there are many degenerated ground states in frustrated system,
the residual entropy is very large and plays an important role in
ordering phenomena.
Number of researchers have studied the residual entropy of triangular
Ising antiferromagnets analytically and they concluded this value is
$0.323k_{\rm B}$ per spin, where the value is about 46.6\% of total
entropy~\cite{Husimi-1950,Syozi-1950,Wannier-1950,Houtappel-1964,Wannier-1973}.

We consider the ground state properties of triangular antiferromagnetic
Ising system with transverse field. The Hamiltonian of this system is given by
\begin{eqnarray}
 \Ham = J \sum_{\left\langle i,j \right\rangle}
  \sigma_i^z \sigma_j^z
  -\Gamma \sum_i \sigma_i^x,
\end{eqnarray}
where $\left\langle i,j \right\rangle$ denotes the pairs of the nearest
neighbor of triangular lattice and $\sigma_i^\gamma$ represents the $\gamma$-component of the Pauli matrix at $i$-th site:
\begin{eqnarray}
 \sigma_i^z = 
  \left(
   \begin{array}{cc}
    1 & 0 \\
    0 & -1
   \end{array}
  \right),
\,\,\,\,\,
\sigma_i^x =
\left(
 \begin{array}{cc}
  0 & 1\\
  1 & 0
 \end{array}
\right).
\end{eqnarray}

\begin{figure}[h]
 \begin{center}
  \includegraphics[scale=1]{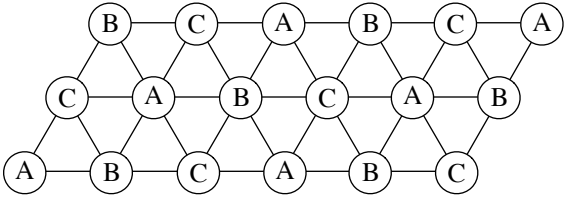}
  \caption{Antiferromagnetic Ising model on triangular lattice. The
  characters in circles A, B, and C indicate the label of sublattice.}
  \label{Fig:tri-n18}
 \end{center}
\end{figure}

\begin{figure}[h]
 \begin{center}
  \includegraphics[scale=0.75]{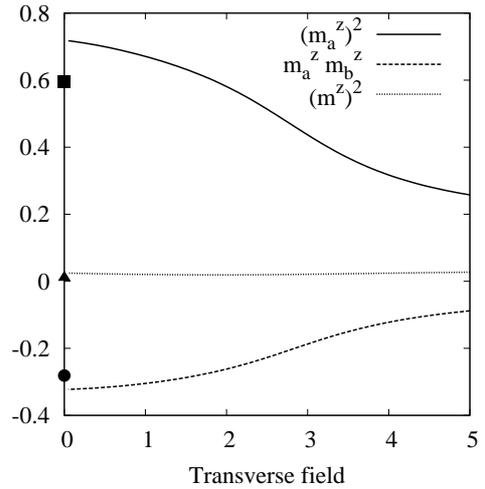}
  \caption{Three lines denote the square of sublattice magnetization 
  $\left( m_{\rm A}^z \right)^2$, the correlation function of sublattice
  magnetization $m_{\rm A}^z m_{\rm B}^z$, and the square of total
  magnetization $\left( m^z \right)^2$ at the ground state as a function of
  transverse field.
  The square, the circle, and the triangle denote $\left( m_{\rm A}^z
  \right)^2$, $m_{\rm A}^z m_{\rm B}^z$, and 
  $\left( m^z \right)^2$ at zero transverse field, respectively.
  The sublattice magnetization at small transverse field is larger than
  that at zero transverse field.
  $\left( m_{\rm A}^z \right)^2 
  = \left( m_{\rm B}^z \right)^2 
  = \left( m_{\rm C}^z \right)^2$ and 
  $m_{\rm A}^z m_{\rm B}^z
  = m_{\rm A}^z m_{\rm C}^z
  = m_{\rm B}^z m_{\rm C}^z$ are satisfied from the symmetry.
  }
  \label{Fig:tridata}
 \end{center}
\end{figure}

Although long-range order does not exist in antiferromagnetic Ising spin
system on triangular lattice at zero temperature and at zero quantum fluctuation, the correlation
function shows power law decay~\cite{Stephenson-1970}:
\begin{eqnarray}
 \left\langle \sigma_i^z \sigma_j^z \right\rangle \propto
  \left| {\bf r}_i - {\bf r}_j \right|^{-1/2},
\end{eqnarray}
where ${\bf r}_i$ represents the position of the $i$-th site.
From this result, we expect that the sublattice order can develop by
additional perturbation such as transverse field.
We consider the correlation function at the ground state of $N=18$
triangular antiferromanget with periodic boundary condition as shown in Figure~\ref{Fig:tri-n18}.
Three sublattices are defined as shown in Figure~\ref{Fig:tri-n18} and the sublattice
magnetization $m_\alpha^z$ and $z$-component of total magnetization
$m^z$ are given by
\begin{eqnarray}
 m_\alpha^z = \frac{3}{N} \sum_{i \in \alpha} \sigma_i^z,
  \,\,\,\,\,
  m^z = \frac{1}{N} \sum_{i} \sigma_i^z
\end{eqnarray}
where $\alpha =$ A, B, and C indicates the label of sublattice.
Figure~\ref{Fig:tridata} shows the sublattice magnetization and the correlation of
sublattice magnetization as a function of transverse field.
$\left( m_{\rm A}^z \right)^2 
= \left( m_{\rm B}^z \right)^2 
= \left( m_{\rm C}^z \right)^2$ 
and 
$m_{\rm A}^z m_{\rm B}^z 
= m_{\rm A}^z m_{\rm C}^z 
= m_{\rm B}^z m_{\rm C}^z$ 
are satisfied from the symmetry.
From Figure~\ref{Fig:tridata}, we can observe the enhancement of the
three-sublattice magnetization at small transverse field.
This result is the example of quantum fluctuation induced ordering~\cite{Moessner-2000,Matsuda-2009}.

\subsection{Decorated Bond System}
\label{}

\begin{figure}[h]
 \begin{center}
  \includegraphics[scale=1]{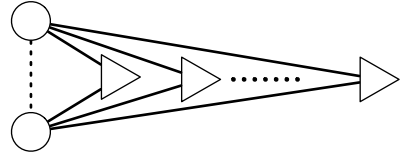}
  \caption{The circles and the triangles denote the system spins and the
  decoration spins, respectively.
  The solid lines and the dotted line denote ferromagnetic coupling $-J$
  and antiferromagnetic coupling $\frac{\Nd}{2}J$, respectively, where
  $\Nd$ is the number of the decoration spins.}
  \label{Fig:dec}
 \end{center}
\end{figure}

We study the correlation function of decorated bond system which is
depicted in Figure~\ref{Fig:dec}.
The circles and the triangles in Figure~\ref{Fig:dec} denote system spins,
$\sigma_1$, $\sigma_2$, and
decoration spins, $s_i$, respectively.
The system spins connect many decoration spins with ferromagnetic
coupling $-J$ depicted solid lines and the system spins connect directly with
antiferromagnetic coupling $\frac{\Nd}{2} J$ depicted dotted line in Figure~\ref{Fig:dec}.
The Hamiltonian of this system is given by
\begin{eqnarray}
 \nonumber
 \Ham =&& -J \sum_i^{\Nd} \left( \sigma_1^z + \sigma_2^z \right) s_i^z
  + \frac{\Nd J}{2} \sigma_1^z \sigma_2^z\\
  &&-\Gamma \left( \sigma_1^x + \sigma_2^x + \sum_i^{\Nd} s_i^x \right),
\end{eqnarray}
where $\Nd$ is the number of the decoration spins.

\begin{figure}[h]
 \begin{center}
  \includegraphics[scale=0.75]{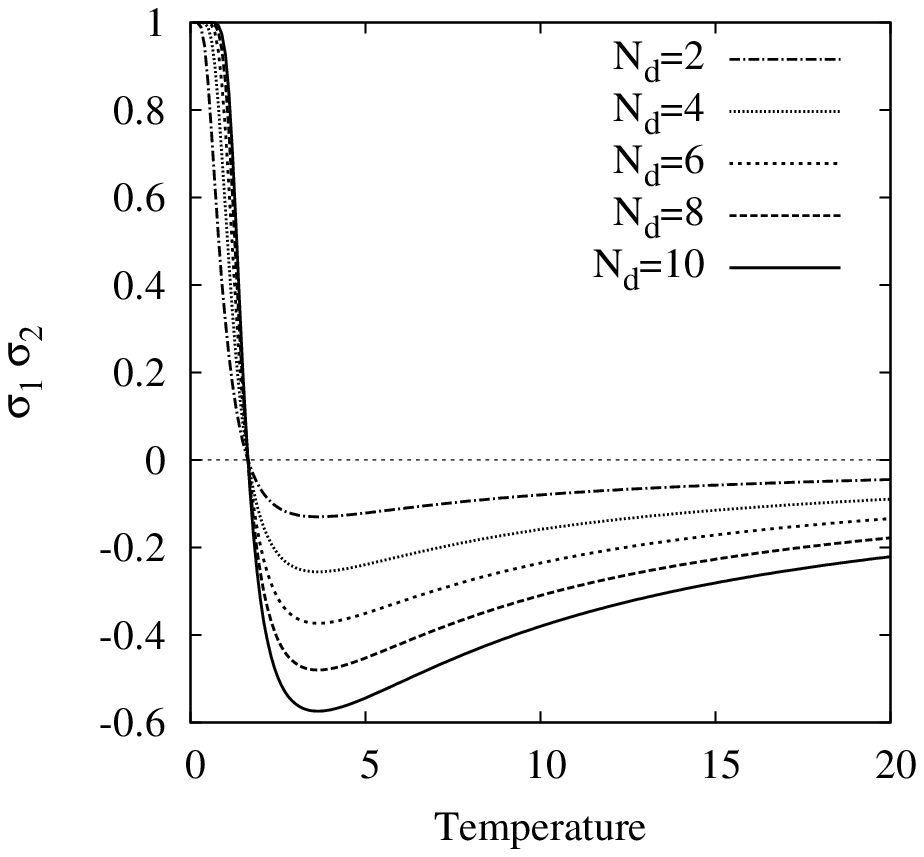}
  \includegraphics[scale=0.75]{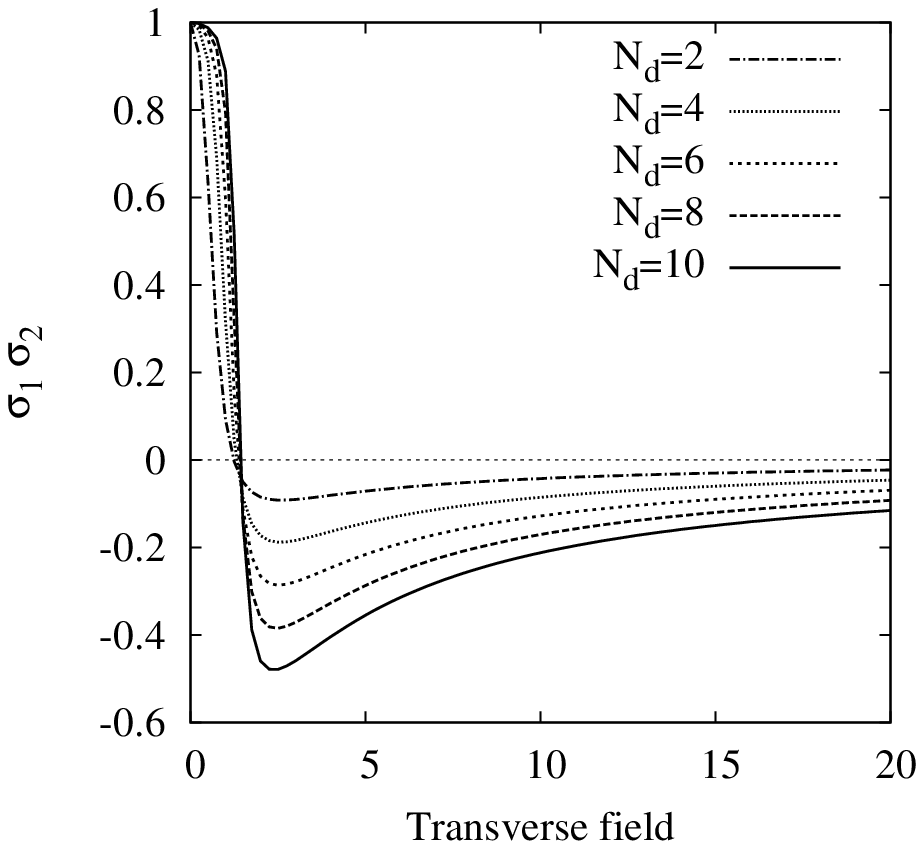}
  \caption{(Upper panel) The correlation function of the system spins as
  a function of temperature.
  (lower panel) The correlation function of the system spins at the ground state as a
  function of transverse field.
  The correlation functions as a function of temperature and transverse
  field show non-monotonic behavior.
  The larger the the number of the decoration spins, the larger the
  absolute value of correlation function.}
  \label{Fig:reentdata}
 \end{center}
\end{figure}

First we consider the case of zero transverse field.
The correlation function between the system spins behaves non-monotonic as a
function of temperature due to entropy effect of decorated spins~\cite{Syozi-1972,Miyashita-1983,Kitatani-1986}.
At $T=0$, all the spins align the same direction, because it is the most favorable state energetically. At finite temperature the decoration spins can flip due to the thermal fluctuation.  When each decoration spins have $\pm 1$ values randomly, the ferromagnetic paths depicted by the solid line in Fig.~\ref{Fig:dec} are weakened, and the direct antiferromagnetic interaction becomes dominant.  The states, in which the decoration spins align randomly, are entropically favorable. 
We can calculate exactly the correlation function between the system
spins by tracing out the degree of the freedom of the decoration spins.
The correlation function is given by
\begin{eqnarray}
 \left\langle \sigma_1^z \sigma_2^z \right\rangle &=& \tanh K_{\rm eff}, \\
K_{\rm eff} &=& \frac{\Nd}{2} 
\log \left( \cosh 2 \beta J \right) - \frac{\beta \Nd J}{2}.
\end{eqnarray}
The upper panel of Figure~\ref{Fig:reentdata} shows the correlation function as a function of temperature.
This is thermal fluctuation induced reentrant behavior~\cite{Syozi-1972,Kitatani-1986}.

Next we consider the case of zero temperature with finite transverse field.
The correlation function between system spins also behaves non-monotonic
as a function of transverse field as shown in the lower panel of Figure~\ref{Fig:reentdata}.
This is the similarity between the thermal fluctuation and the quantum fluctuation.
From this result, we expect that the lattice systems with decorated
bonds show quantum reentrant phase transition as well as thermal
reentrant phase transition~\cite{Tanaka-2009-2}.

\section{Conclusion and Future Works}
\label{sec:conclusion}

In this paper, we considered the ordering in homogeneous frustrated
systems due to quantum fluctuation comparing with thermal fluctuation.
In antiferromagnetic Ising spin system on triangular lattice, we observed enhancement
of three-sublattice spin structure at finite transverse field.
We also studied the quantum reentrant behavior of the correlation function of the
decorated bond systems as well as thermal fluctuation.
In this study, we focus on only the ground state static properties.
It is future work to clarify the dynamical ordering process, which is related to
the quantum
annealing~\cite{Matsuda-2009,Finnila-1994,Kadowaki-1998,Farhi-2001,Das-2005,Santoro-2006,Tanaka-2007-2,Das-2008,Tanaka-2009-1,Kurihara-2009,Sato-2009,Tanaka-2010-1,Tanaka-2010-2}.

\section*{Acknowledgment}

This work was partially supported by Research on Priority Areas 
``Physics of new quantum phases in superclean materials'' (Grant
No. 17071011) from MEXT and 
by the Next Generation Super Computer Project, Nanoscience Program from MEXT.
S.T. is partly supported by Grant-in-Aid for Young Scientists Start-up
(No.21840021) from JSPS.
The authors also thank the Supercomputer Center, Institute for Solid State Physics, University 
of Tokyo for the use of the facilities. 

%% The Appendices part is started with the command \appendix;
%% appendix sections are then done as normal sections
%% \appendix

%% \section{}
%% \label{}

%% References
%%
%% Following citation commands can be used in the body text:
%% Usage of \cite is as follows:
%%   \cite{key}         ==>>  [#]
%%   \cite[chap. 2]{key} ==>> [#, chap. 2]
%%

%% References with bibTeX database:

%\bibliographystyle{elsarticle-num}
%\bibliography{<your-bib-database>}

%% Authors are advised to submit their bibtex database files. They are
%% requested to list a bibtex style file in the manuscript if they do
%% not want to use elsarticle-num.bst.

%% References without bibTeX database:

% \begin{thebibliography}{00}

%% \bibitem must have the following form:
%%   \bibitem{key}...
%%

% \bibitem{}

% \end{thebibliography}

\end{document}